\documentclass[aps,prl,a4paper,twocolumn]{revtex4}

\usepackage{graphics}
\usepackage{amssymb}
\usepackage{amsmath}
\usepackage{hyperref}
\usepackage{xcolor}
\usepackage{physics}
\usepackage{graphicx}
\usepackage{fancyhdr}
\usepackage{bm}
\usepackage{float}
\usepackage{braket}
\usepackage{svg}
\usepackage{times}
\usepackage{courier}
\usepackage{bm}
\usepackage{subfig}
\usepackage{mdframed}
\usepackage{dsfont}
\usepackage{tikz}
\usetikzlibrary{shapes,arrows}
\usepackage{bbm}
\usepackage{verbatim}
\usepackage{soul}
\usepackage{indentfirst}
\hypersetup{
    colorlinks=true,
    linkcolor=blue,
    urlcolor=cyan,
    citecolor=cyan,}

\usetikzlibrary{decorations.markings}

\newcommand{\id}{\mathds{1}}
\definecolor{maroon}{RGB}{128, 0, 0}

\newcommand\bi{\begin{itemize}}
\newcommand\ei{\end{itemize}}

\newcommand\cV{{\cal V}}

\newcommand\cD{{\cal D}}
\newcommand\cC{{\cal C}}
\newcommand\cZ{{\cal Z}}
\newcommand\cS{{\cal S}}

\newcommand{\cO}{{\cal O}}

\newcommand{\cA}{{\cal A}}
\newcommand{\cH}{{\cal H}}

\newcommand\ZZ{\hbox{Z\kern-.4emZ}}
\newcommand\sZZ{\hbox{\sevenfont Z\kern-.4emZ}}

\newcommand{\cM}{{\cal M}}

\newcommand{\Comment}[1]{{}}

\begin{document}


\title{Generalized Symmetry Resolution of Entanglement in CFT for Twisted and Anyonic sectors}

\author{A. Das$^{(1), (2)}$}
\email{arpit.das@ed.ac.uk}

\author{J. Molina-Vilaplana$^{(3)}$}
\email{javi.molina@upct.es}

\author{P. Saura-Bastida$^{(3)}$}
\email{pablo.saura@upct.es}

\affiliation{${}^{(1)}$School of Maths, University of Edinburgh, Edinburgh, EH9 3FD, U.K.}
\affiliation{${}^{(2)}$Higgs Centre for Theoretical Physics, University of Edinburgh, Edinburgh EH8 9YL, U.K.}
\affiliation{${}^{(3)}$Universidad Polit\'ecnica de Cartagena. C/Dr Fleming S/N. 30202, Cartagena, Spain.}


\begin{abstract}

A comprehensive symmetry resolution of the entanglement entropy (EE) in $(1+1)$-d rational conformal field theories (RCFT) with categorical non-invertible symmetries is presented. This amounts to symmetry resolving the entanglement with respect to the generalized twisted and anyonic charge sectors of the theory. The anyonic sectors label the irreducible representations of a modular fusion category defining the symmetry and can be understood through the $(2+1)$-d symmetry topological field theory (SymTFT) that encodes the symmetry features of the CFT. Using this, we define the corresponding generalized boundary dependent charged moments necessary for the symmetry resolution of the entanglement entropy, which is the main result of this work. Furthermore, contrary to the case of invertible symmetries, we observe the breakdown of entanglement equipartition between different charged sectors at the next-to-leading order in the ultraviolet cutoff.
\end{abstract}

\maketitle


\textit{Introduction}\textemdash
Symmetry resolved entanglement entropy (SREE) is an entanglement measure, that, in extended quantum systems such as quantum field theories (QFTs) with an internal global symmetry,  quantifies the amount of entanglement entropy (EE) corresponding to different symmetry sectors. Its study has fostered interest in symmetries and their role in shaping the entanglement structure of these theories (see \cite{Castro-Alvaredo:2024azg} for a short review on the topic).
   
In this sense, the concept of a global symmetry in QFT  has been recently generalized with respect to the standard ways given by group theory. In the new paradigm, one realizes that any symmetry operation can be associated with a topological operator \cite{Gaiotto:2014kfa}. These generalizations correspond to \emph{higher-form} symmetries,  related to the conservation of extended objects, and \emph{categorical} or non-invertible symmetries, for which, the composition law of their topological operators is given by a \emph{fusion category}, in contrast to the simple group multiplication law related to the standard group-like \emph{invertible} symmetries (a pedagogical review can be found in  \cite{McGreevy:2022oyu}).
  
The aim of this work is to obtain a full symmetry resolution of the entanglement entropy for $(1+1)$-d RCFT with a categorical non-invertible symmetry and, thus, to generalize the proposal in \cite{Saura-Bastida:2024yye}. Furthermore, a very nice relation between asymptotic symmetry resolved density of states (ASRDoS) as explored in \cite{Lin:2022dhv} and SREE via boundary CFT (BCFT) was alluded to in Table 1 of \cite{Kusuki:2023bsp} and established for the group-like case therein. \cite{Saura-Bastida:2024yye} extended this relation to the case of categorical symmetries relevant to the untwisted sector. Another aim of this work is to extend this nice relation between ASRDoS and SREE for categorical symmetries relevant to the twisted and anyonic sectors.

Central to our developments is to have suitable representations and, therefore, suitable notions of generalized charged sectors, for these categorical symmetries. Remarkably, it has been realized that the symmetry features of a $(1+1)$-d QFT with a non-invertible symmetry defined by a modular fusion category (MFC) $\cC$ are encoded by a $(2+1)$-d topological quantum field theory coupled to it, recently referred to as symmetry TFT (SymTFT) \cite{Turaev:1992hq, Freed:2022qnc}. This construction is especially useful in understanding what a ``generalized charge sector'' means in the case of categorical non-invertible symmetries, allowing to properly define partition functions in such charged sectors, which is one of the main results of this work. 

Indeed, representations of generalized global symmetries have been recently studied from the perspective of SymTFT in \cite{Bhardwaj:2023ayw}. While this observation on a  physical context for $(1+1)$-d CFTs was firstly noted in \cite{Lin:2022dhv}, authors in  \cite{Bhardwaj:2023ayw} find that a similar structure is also valid in the case of $(2+1)$-d theories with fusion 2-category symmetries (see also \cite{Bartsch:2023wvv}). 

Our results shed light on the relationship between, non-invertible symmetries, boundary conditions and quantum correlations, pointing to the fact that an extension of the program to higher dimensions, could provide powerful tools to study topological order and phases.

\textit{Rational CFT basics}\textemdash
An RCFT is a 2D CFT with a finite number of primary fields: $a$ which are the highest weight integral irreps w.r.t. its symmetry/chiral algebra $\cA$ \footnote{For simplicity we shall consider diagonal RCFT where the torus partition function is a diagonal modular invariant \cite{DiFrancesco:1997nk}}. These irreps arrange themselves as \textit{Verma modules}: $\cV_a$ where its levels contain primaries and their descendants. With this, the bulk Hilbert space decomposes as \cite{DiFrancesco:1997nk, Chang:2018iay, Lin:2022dhv}:
\begin{align}
    \cH_\mathds{1} = \bigoplus_{a\in\cC} \cV_a\otimes\overline{\cV_a}, \label{H1}
\end{align}

\Comment{
\raisebox{-11pt}{
\begin{tikzpicture}
\draw [dashed, decoration = {markings, mark=at position 0.6 with {\arrow[scale=1]{stealth}}}, postaction=decorate] (-1,0.5) -- node[below]{$_\mathds{1}$} (1,0.5);
\draw[draw=blue!60, line width=0.5mm] (-1,0) -- (-1,1); 
\draw[draw=blue!60, line width=0.5mm] (1,0) -- (1,1); 
\draw[draw=brown!60, line width=0.5mm] (-1,1) -- (1,1); 
\draw[draw=brown!60, line width=0.5mm] (-1,0) -- (1,0); 
\end{tikzpicture}
}
}

The torus partition function of the theory is given as:
\begin{align}
    \cZ(q,\bar{q}) &= \tr_{\cH_\mathds{1}}\left(q^{L_0-\frac{c}{24}}\bar{q}^{\bar{L}_0-\frac{\bar{c}}{24}}\right) = \raisebox{-11pt}{
\begin{tikzpicture}
\draw [dashed, decoration = {markings, mark=at position 0.6 with {\arrow[scale=1]{stealth}}}, postaction=decorate] (-1,0.5) -- node[below]{$_\mathds{1}$} (1,0.5) ;
\draw (-1,0) rectangle (1,1);
\end{tikzpicture}
} \nonumber\\
&=\sum_{a}\chi_a(q)\overline{\chi_a(q)}, \label{partn_fn}
\end{align}
where $\mathds{1}$ in the subscript of $\cH$ denotes that the bulk Hilbert space is the untwisted Hilbert space or trivially twisted by the identity {\it topological defect line} (TDL), $\id$ (more on this below). $\cH_\id$ is defined on a time slice along spatial $S^1$, ($\bar{L}_0$) $L_0$ are the (anti-)holomorphic Virasoro zero-modes and ($\bar{c}$) $c$ are the (anti-)holomorphic central charges. $q=e^{2\pi\text{i}\tau}$ where $\tau$ is the modular parameter on the torus and $\chi_a(q)$ denotes the linearly independent characters \cite{Das:2020wsi}. The no. of characters is less than or equal to the no. of primaries \cite{DiFrancesco:1997nk, Das:2023qns}. The black box in \eqref{partn_fn} pictorially represents the torus partition function with the temporal $S^1$ along the vertical axis and the spatial $S^1$ along the horizontal axis. 

Note that, the torus is diffeomorphic under the action of the modular group $\gamma\in\text{SL}_2(\mathbb{Z})=\langle S,T\rangle$. So, another important ingredient of an RCFT is its \textit{modular data}: $S:\tau\to-\frac{1}{\tau}$ and $T:\tau\to\tau+1$ matrices \cite{Harvey:2018rdc}. Additionally, one also needs to specify what are called the OPE coefficients which dictate how primaries behave when they are brought close to each other inside correlation functions. 

Roughly, the above data also define what is called a modular fusion category (MFC): $\cC$. Its elements (or simple objects): $a,b,c,d,\ldots \in\mathcal{C}$ are in one-to-one correspondence with the primaries. Furthermore, their \textit{fusion rule/algebra}: $a \times d = \sum_c N^c_{ad} \, \, c \,,$ gives rise to the OPE coefficients. From these, one can obtain the $S$ and $T$ matrices \cite{rowell2009classification}. This is how an MFC encodes the modular data of an RCFT. The fusion coefficients -- in the fusion rule -- $N^c_{ad}\in\mathbb{Z}_{\geq 0}$ are given by the Verlinde formula: $N^c_{ad} = \sum\limits_{b}\frac{S_{ba}S_{bd}S^{*}_{bc}}{S_{1b}}$, where $S_{ab}$ denotes an element of the $S$-matrix of the RCFT, with the subscript `$1$' labelling the vacuum of the theory: $|1\rangle$. 

It is known that the action of TDLs generate 0-form global symmetries in 2D RCFTs \cite{Chang:2018iay}. Note that we are considering a diagonal RCFT. So its topological defect lines (TDLs): $b$, are in on-to-one correspondence with its bulk primaries -- hence the same label for both. Furthermore, in this scenario these TDLs are referred to as Verlinde lines and they commute with all the generators of the chiral algebra. Additionally, they satisfy the same fusion algebra as that of the bulk primaries. Hence, in general this suggests that lines $a$ are non-invertible in nature as, in general, their fusion is non-group like acting on the bulk primaries as \cite{Chang:2018iay, Lin:2022dhv, Hegde:2021sdm},
\begin{align}
    b |\phi_a\rangle = \frac{S_{ba}}{S_{1a}}|\phi_a\rangle. \label{tdl_local}
\end{align}
\textit{Torus partition function and TDL insertions}\textemdash
Consider inserting a TDL $b$, along the spatial $S^1$. This amounts to it acting on the bulk primaries as in \eqref{tdl_local}. Then, we have from \eqref{partn_fn} \cite{Chang:2018iay, Lin:2022dhv}:
\begin{align}
    \cZ^{b}(q,\bar{q}) &= \tr_{\cH_\mathds{1}}\left(b \, q^{L_0-\frac{c}{24}}\bar{q}^{\bar{L}_0-\frac{\bar{c}}{24}}\right) = \raisebox{-11pt}{
\begin{tikzpicture}
\draw [draw=maroon, thick, decoration = {markings, mark=at position 0.6 with {\arrow[scale=1]{stealth}}}, postaction=decorate] (-1,0.5) -- node[below] {$_b$} (1,0.5);
\draw (-1,0) rectangle (1,1);
\end{tikzpicture}
}\nonumber\\
&=\sum_{a}\frac{S_{ba}}{S_{1a}}\chi_a(q)\overline{\chi_a(q)}. \label{partn_fn_spatial}
\end{align}
We can also consider inserting a TDL along the temporal $S^1$. In this case, the Hilbert space will be modified to give rise to a {\it defect} Hilbert space twisted by the TDL. Then, the twisted partition function is given as \cite{Hegde:2021sdm}:
\begin{align}
    &\cZ_{a}(q,\bar{q}) = \tr_{\cH_a}\left(q^{L_0-\frac{c}{24}}\bar{q}^{\bar{L}_0-\frac{\bar{c}}{24}}\right) 
    = \raisebox{-12pt}{
    \begin{tikzpicture}
        \draw [draw=maroon, thick, decoration = {markings, mark=at position 0.6 with {\arrow[scale=1]{stealth}}}, postaction=decorate] (0,0) -- node[right]{$_a$} (0,1) ;
        \draw (-1,0) rectangle (1,1);
    \end{tikzpicture}
    }\nonumber\\
    &=\raisebox{-26pt}{
    \begin{tikzpicture}
        \draw [draw=maroon, thick, decoration = {markings, mark=at position 0.6 with {\arrow[scale=1]{stealth}}}, postaction=decorate] (1,1) -- node[above]{$_a$} (0,1) ;
        \draw (0,0) rectangle (1,2);
    \end{tikzpicture}
    }
    = \sum_{b}\frac{S_{\bar{a}b}}{S_{1b}}\chi_b(\tilde{q})\overline{\chi_b(\tilde{q})} = \sum_{b,c,d}\frac{S_{\bar{a}b}}{S_{1b}}S_{bd}S^{*}_{bc}\chi_d(q)\overline{\chi_c(q)} \nonumber\\
    & = \sum_{c,d}N^c_{ad}\chi_d(q)\overline{\chi_c(q)}, \label{partn_fn_temporal}
\end{align}
where in the second line above, we have used a modular $S$-transformation to get the following relation: $\cZ_{a}(q,\bar{q})=\cZ^{a}(\tilde{q},\bar{\tilde{q}})$ with $\tilde{q}=e^{-\frac{2\pi\text{i}}{\tau}}$ where $S:q\to\tilde{q}$. In the last equality above we have used \eqref{S_prop} along with the assumption that the given RCFT has only self-conjugate irreps. This implies $\bar{a}=a$, where $\bar{a}$ is the orientation reversal line of $a$. Hence, the $S$-matrix is real.

So, from \eqref{partn_fn_temporal} we get,
\begin{align}
    \cH_{a} = \bigoplus_{c,d} N^c_{ad} \, \cV_c\otimes\overline{\cV_d}. \label{def_decomp}
\end{align}

\textit{SREE for invertible symmetries}\textemdash
Before proceeding, let us introduce a pictorial notation here. As explained above, a black box would imply a torus partition function -- as in \eqref{partn_fn_spatial}. Spatial TDL insertion will be denoted by a horizontal maroon arrow and temporal TDL insertion by a vertical maroon arrow -- as in \eqref{partn_fn_temporal} and \eqref{eq:proj_Pi_r}. However, a brown box with blue colored right and left edges will denote the annulus partition function (as described in Fig. \ref{conf_W} of Supplemental Material, SM) rather than the torus partition function. The annulus partition function would include the interval $A$ and at its end points the conformal boundary conditions -- labelled by the blue edges -- as in \eqref{asymp_4}.

In an RCFT with a discrete invertible symmetry described by an Abelian group $G$, the bulk Hilbert space $\cH_\id$ decomposes into various subspaces $\cH_r$ with the projectors into these subspaces (or, charged sectors/irreps labelled by $r$), $P_r:\cH_\id\to \cH_r$ given by \cite{Kusuki:2023bsp},
\begin{align} 
\label{eq:proj_Pi_r}
P_r = 
 \frac{d_r}{|G|} \sum_{g \in G} \chi^*_r(g)\, 
\begin{tikzpicture}
\begin{scope}[draw=maroon, thick, decoration={markings, mark=at position 0.5 with {\arrow{>}}}] 
    \draw[postaction={decorate}] (-0.75,-0.1)--(0.75,-0.1) node[pos=0.4,above] {$g$};
\end{scope}
\node[draw=none] at (1.2,0) {$\left.\right.$};
\end{tikzpicture} 
, \forall \, \, g \in G
\end{align}
where $d_r$ is the dimension of the irrep, $|G|\equiv \sum\limits_{r}d_r^2$ is the volume of the group and $\chi_r^*(g)$ is the character of the element of $g \in G$ in the irrep $r$. The spatial TDL $g$ acts on states supported on the region $A$ of the Hilbert space.  

With these projectors, the projected partition function associated to the $r$-charged sector reads as (see \cite{Saura-Bastida:2024yye} and SM)
\begin{align} \label{eq:part_func}
\cZ_{\alpha \beta}[q^n,r] &= \tr\, \left(P_r \, \rho_A^n\right) =\frac{d_r}{|G|} \sum_{g \in G} \chi^*_r(g) \frac{Z_{\alpha \beta}[q^n,g]}{\left(Z_{\alpha \beta}[q]\right)^n}\, ,\\ \nonumber
Z_{\alpha \beta}[q^n,g] &=  \tr\left[g \, q^{n\left(L_0-c/24\right)}\right]\, ,
\end{align}
where $\rho_A$ denotes the reduced density matrix for the region $A$. In the above definitions all the traces (over {\it open string} states) are subjected to explicit boundary conditions $\alpha$ and $\beta$ on the entangling points of the interval $A$ and the action of TDL $g$ is encoded in the \emph{charged moments} $Z_{\alpha \beta}[q^n,g]$ -- defined above in the \emph{open string} channel (see Fig. \ref{conf_W} in SM).

Using the BCFT approach to EE (see SM), one may express $Z_{\alpha \beta}[q^n,g]$ in the $S$-dual channel (or, in the \emph{closed string} channel) in terms of  boundary states $\widetilde{\ket{\alpha}}_g$ and $\widetilde{\ket{\beta}}_g$ as: $Z_{\alpha \beta}[q^n,g] =   {}_g\widetilde{{\langle \alpha}|} \tilde{q}^{\frac{1}{n}\left(L_0 - c/24\right)} \widetilde{|\beta\rangle}_g\, ,$
where the sub-index $g$ refers to the fact that states $\widetilde{\ket{\alpha}}_g$ and $\widetilde{\ket{\beta}}_g$ are conformal Cardy states belonging to the \emph{defect} or \emph{twisted} Hilbert space $\mathcal{H}_{g}$ -- twisted by $g$ \cite{Cardy:2004hm, Kusuki:2023bsp}. We note that as $Z_{\alpha \beta}[q^n,g]$ is defined through the insertion of $g$ in the annulus partition function (in the \emph{open string} channel), it is required that $g$ can end {\it topologically} on the boundary of the interval since it is a symmetry defect \cite{Kusuki:2023bsp}. This imposes a constraint on the allowed boundary states in the dual $S$-channel (or, in the \emph{closed string} channel) \cite{Choi:2023xjw}. For invertible group-like symmetries the topological endability is equivalent to having $G$-invariant boundary states such that $g'\widetilde{\ket{\alpha}}_e =\widetilde{\ket{\alpha}}_e \,,~~~\forall ~g'\in G$ and where $e$ denotes the identity element of the group.

With this, it has been shown \cite{Kusuki:2023bsp,Casini:2019kex,Magan:2021myk} that the SREE defined by \cite{Goldstein:2017bua},
\begin{align}
    &\cS_n[q,r] = \frac{1}{1-n} \log\frac{\cZ_{\alpha \beta}[q^n,r]}{(\cZ_{\alpha \beta}[q,r])^n}, \, \, \, \cS[q,r] = \lim_{n\to 1}\cS_{n}[q,r], \label{srre_ee}
\end{align}
in the $\tilde{q}\to 0\, (\ell \gg \epsilon)$ limit reads as
\begin{align}
    &\cS[q,r] = \frac{c}{3}\log\frac{\ell}{\epsilon} + (g_\alpha^{*} + g_\beta) + \log\frac{d_r^2}{|G|} + \cO\left(\frac{\epsilon}{\ell}\right), \label{srre_00}
\end{align}
where $g_\alpha\equiv \log\langle 1\widetilde{|\alpha\rangle}$ is the {\it Affleck-Ludwig} boundary entropy. Comparing \eqref{srre_00} with \eqref{EE_2} in SM, we see that at leading order in the UV-cutoff, the entanglement entropy of region $A$ associated to the $r$-charged sector $S[q,r]$ is equal for all sectors, that is, independent of $r$. This is known  as  {\it entanglement equipartition} \cite{Xavier2018, Northe:2023khz, Kusuki:2023bsp}. Indeed for Abelian symmetry group $G$, the equipartition holds even at the sub-leading order: $\cO(1)$ while for non-Abelian symmetry group it breaks at this order (see \cite{Goldstein:2017bua, Calabrese:2021wvi, Milekhin:2021lmq, DiGiulio:2022jjd, Kusuki:2023bsp}). Note that, for group-like symmetries generated by the action of Verlinde lines, we would always have an Abelian group since the commutativity of fusion rules implies that the symmetry group is Abelian.

\textit{Categorical-SREE}\textemdash
In \cite{Saura-Bastida:2024yye}, the above group-like symmetry resolution was extended to the case of non-invertible symmetries by considering categories involving non-invertible lines.

While considering the action of Verlinde lines on primaries, one needs to distinguish their action on bulk primaries and on defect primaries. TDLs map bulk primaries to bulk primaries, that is, $b:\cH_\id\to\cH_\id$ owing to \eqref{tdl_local}. However, their action on defect primaries is given by the map: $b:\cH_a\to\cH_c$, that is, twisted sectors (or, defect spaces) are mapped into each other. In \cite{Saura-Bastida:2024yye}, categorical symmetry resolution was performed w.r.t. the Verlinde lines acting on the untwisted sector or $\cH_\id$. In this spirit, we can term this symmetry resolution as \emph{untwisted} CaT-SREE. The two main ingredients needed to do this are the relevant projectors, as in \eqref{eq:proj_Pi_r} and the notion of topological endability of non-invertible lines. The latter concept was introduced in \cite{Choi:2023xjw} from which we only need the notion of $\cC$-weakly symmetric boundary states to define CaT-SREE as was advocated in \cite{Saura-Bastida:2024yye}. Given a category $\cC$, a conformal boundary state is $\cC$-weakly symmetric if: $a \widetilde{|\beta\rangle} = \widetilde{|\beta\rangle} + \ldots, \quad \forall \, \, a\in\cC$. If weakly symmetric states exist, we can symmetry resolve w.r.t. the given category (more details can be found in SM). 

Additionally, we make an important observation: at least one non-invertible line needs to exist in an MFC in order to perform symmetry resolution of entanglement w.r.t. the action of Verlinde lines -- either for the group-like case or the categorical case (see SM for a simple argument).

\textit{Twisted CaT-SREE}\textemdash
In this section we generalise \cite{Saura-Bastida:2024yye} by allowing for the action of TDLs on defect primaries. Let us study the spectrum (or, operator content) of the defect spaces. The spectrum can be computed using \eqref{partn_fn_temporal} which results in the decomposition as given in \eqref{def_decomp} (see SM). From this, one needs to construct relevant projectors which map from $\cH_{a\neq\mathds{1}}$ to the subspace: $\cV_c\otimes\overline{\cV_d}, \ \ \forall \, a,c,d\in\cC$. Following \cite{Lin:2022dhv}, we first define, the following two sets of orthogonal projectors:
\begin{align}
&P_a^{(c,\ast)}\equiv \sum_{b}S_{1c}S^*_{bc}
\raisebox{-14pt}{
\begin{tikzpicture}
\draw [draw=maroon, thick, decoration = {markings, mark=at position 0.8 with {\arrow[scale=1]{stealth}}}, postaction=decorate] (-1,0) -- (1,0) node[right]{$b$};
\draw [preaction={draw=white,line width=6pt}, draw=maroon, thick, decoration = {markings, mark=at position 0.9 with {\arrow[scale=1]{stealth}}}, postaction=decorate] (0,-0.5) -- (0,0.5) node[above]{$a$};
\end{tikzpicture}
}, \label{MFC.projector_cd} \\
&P_a^{(\ast,d)}\equiv \sum_{b}S_{1d}S^*_{bd}
\raisebox{-12pt}{
\begin{tikzpicture}
\draw [draw=maroon, thick, decoration = {markings, mark=at position 0.9 with {\arrow[scale=1]{stealth}}}, postaction=decorate] (0,-0.5) -- (0,0.5) node[above]{$a$};
\draw [preaction={draw=white,line width=6pt}, draw=maroon, thick, decoration = {markings, mark=at position 0.8 with {\arrow[scale=1]{stealth}}}, postaction=decorate] (-1,0) -- (1,0) node[right]{$b$};
\end{tikzpicture}
}. \label{MFC.projector_cd2}
\end{align}
It can be shown using the definitions above \eqref{MFC.projector_cd}-\eqref{MFC.projector_cd2}, that the sets $\left\{{P_{a}}^{(c,\ast)}\right\}_{a\in\cC}$ and $\left\{{P_a}^{(\ast,d)}\right\}_{a\in\cC}$ are individually two complete set of projectors with each set being individually idempotent too \cite{Lin:2022dhv}.

Now, from these two sets, let us define the relevant projectors as: $P_a^{(c,d)}\equiv P_a^{(c,\ast)}P_a^{(\ast,d)}$ -- whose action is given as: $P_a^{(c,d)}:\cH_a\to\cV_c\otimes\overline{\cV_d}$ \cite{Lin:2022dhv}. Note that, when $a=\mathds{1}$ then $c=d$ due to the decomposition given in \eqref{H1}.

From above, first let us define the respective projected torus partition functions as \cite{Lin:2022dhv},
\begin{align}
\tr \left[P_a^{(c,d)} \rho\right] 
&\equiv \tr_{\cH_a^{(c,d)}}\, q^{L_0 - c/24} \, \bar{q}^{\bar{L}_0-\frac{\bar{c}}{24}} =
\raisebox{-20pt}{
\begin{tikzpicture}[scale=.9]
\draw [draw=maroon, thick, decoration = {markings, mark=at position 0.8 with {\arrow[scale=1]{stealth}}}, postaction=decorate] (-1,1) -- (1,1) node[right]{$c$};
\draw [preaction={draw=white,line width=6pt}, draw=maroon, thick, decoration = {markings, mark=at position 0.9 with {\arrow[scale=1]{stealth}}}, postaction=decorate] (0,0) -- (0,1.5) node[above]{$a$};
\draw [preaction={draw=white,line width=6pt}, draw=maroon, thick, decoration = {markings, mark=at position 0.8 with {\arrow[scale=1]{stealth}}}, postaction=decorate] (-1,0.5) -- (1,0.5) node[right]{$d$};
\draw (-1,0) rectangle (1,1.5);
\end{tikzpicture}
}\nonumber\\ 
&=\sum_{b,b^\prime}S_{1c}S^*_{bc}S_{1d}S^*_{b^\prime d}
\raisebox{-20pt}{
\begin{tikzpicture}[scale=.9]
\draw [draw=maroon, thick, decoration = {markings, mark=at position 0.8 with {\arrow[scale=1]{stealth}}}, postaction=decorate] (-1,1) -- (1,1) node[right]{$b$};
\draw [preaction={draw=white,line width=6pt}, draw=maroon, thick, decoration = {markings, mark=at position 0.9 with {\arrow[scale=1]{stealth}}}, postaction=decorate] (0,0) -- (0,1.5) node[above]{$a$};
\draw [preaction={draw=white,line width=6pt}, draw=maroon, thick, decoration = {markings, mark=at position 0.8 with {\arrow[scale=1]{stealth}}}, postaction=decorate] (-1,0.5) -- (1,0.5) node[right]{$b^\prime$};
\draw (-1,0) rectangle (1,1.5);
\end{tikzpicture}
}. \label{eval_F}
\end{align}
where $\rho$ above denotes the full density matrix of the theory \footnote{Note that, when addressing the EE of twisted/anyonic sectors of a theory defined in a manifold without boundaries, the fusion rules of the TDLs are those defined in the bulk. Contrarily, in theories that are explicitly defined on manifolds with boundaries, a deformed version of the bulk fusion rules must be applied \cite{Choi:2024tri, Choi:2024wfm, Bhardwaj:2024igy, Heymann:2024vvf, GarciaEtxebarria:2024jfv}.}. 

Now $b$ and $b^\prime$ above can be fused and this calculation is present in the SM (see \cite{Lin:2022dhv} for more details). Then from \eqref{asymp_3} we get,
\begin{align}
    \tr \left[P_a^{(c,d)} \rho\right] & \sim \, \frac{N^c_{ad}}{|\cC|} \, \frac{d_c \, d_d}{d_a} \,
    \vcenter{\hbox{\begin{tikzpicture}
        \draw (-.5,-1) rectangle (.5,1);
        \draw [draw=maroon, thick, decoration = {markings, mark=at position 0.6 with {\arrow[scale=1]{stealth}}}, postaction=decorate]  (0,-1) -- node(A)[right] {$_a$} (0,1);
        \end{tikzpicture}}} \nonumber\\ 
        &\overset{\text{under }S^{-1}}{=} \, \, \, \frac{N^c_{ad}}{|\cC|} \, \frac{d_c \, d_d}{d_a} \,
        \vcenter{\hbox{\begin{tikzpicture}
        \draw [draw=maroon, thick, decoration = {markings, mark=at position 0.6 with {\arrow[scale=1]{stealth}}}, postaction=decorate] (-1,0.5) -- node[below] {$_a$} (1,0.5);
        \draw (-1,0) rectangle (1,1);
        \end{tikzpicture}}} \label{torus_4}
\end{align}
where we have used $|\cC|=\frac{1}{S_{11}^2}$ (see SM for more details). Note that, here, first we are constructing the torus partition function as above and then next from it the annulus partition function will be constructed below by taking the interval $A$ of length $\ell$ in \eqref{torus_4} (see SM for details as to why this is important in the twisted case),
\begin{align}
    \cZ_{\alpha\beta}^a[q,(c,d)] &\equiv \cZ^a[q,(c,d)] \sim \frac{N^c_{ad}}{|\cC|} \, \frac{d_c \, d_d}{d_a} \,
        \vcenter{\hbox{\begin{tikzpicture}
        \draw [draw=maroon, thick, decoration = {markings, mark=at position 0.6 with {\arrow[scale=1]{stealth}}}, postaction=decorate] (-1,0.5) -- node[below] {$_a$} (1,0.5);
        \draw[draw=blue!60, line width=0.5mm] (-1,0) -- (-1,1); 
        \draw[draw=blue!60, line width=0.5mm] (1,0) -- (1,1); 
        \draw[draw=brown!60, line width=0.5mm] (-1,1) -- (1,1); 
        \draw[draw=brown!60, line width=0.5mm] (-1,0) -- (1,0); 
        \end{tikzpicture}}} \nonumber\\
    & \overset{\text{under }S}{=} \, \, \, \frac{N^c_{ad}}{|\cC|} \, \frac{d_c \, d_d}{d_a} \,
    \vcenter{\hbox{\begin{tikzpicture}
        \draw[draw=brown, line width=0.5mm] (-0.5,-1) -- (-0.5,1); 
        \draw[draw=brown, line width=0.5mm] (0.5,-1) -- (0.5,1); 
        \draw[draw=blue, line width=0.5mm] (-0.5,1) -- (0.5,1); 
        \draw[draw=blue, line width=0.5mm] (-0.5,-1) -- (0.5,-1); 
        \draw [draw=maroon, thick, decoration = {markings, mark=at position 0.6 with {\arrow[scale=1]{stealth}}}, postaction=decorate]  (0,-1) -- node(A)[right] {$_a$} (0,1);
        \end{tikzpicture}}}  \nonumber\\
    &= \, \frac{N^c_{ad}}{|\cC|} \, \frac{d_c \, d_d}{d_a}\left({}_a\langle \widetilde{\beta_1|}\tilde{q}^{(L_0-\frac{c}{24})}\widetilde{|\beta_2\rangle}_a\right) \nonumber\\     
    &\overset{\tilde{q} \to 0}{\sim} \, \, \, \frac{N^c_{ad}}{|\cC|} \, \frac{d_c \, d_d}{d_a} \, g^a_{\beta_1} \, \bar{g}^a_{\beta_2} \, \tilde{q}^{h_0^a-\frac{c}{24}}, \label{asymp_4}
\end{align}
where in the first line we have explicitly introduced the weakly symmetric boundary conditions: $\alpha,\beta$ to denote that we are computing the annulus partition functions with these boundary conditions. For simplicity, we shall drop the notation of $\alpha,\beta$ from here on. In the last line above we have used \eqref{g-fn_neq}. Also, $g^a_{\beta_i} \equiv {}_a\widetilde{\langle \beta_i|}1 \rangle_a$ can be referred to as the {\it $a$-twisted boundary entropy}. It computes the overlap of the twisted Cardy state $\widetilde{|\beta_i\rangle}$ with the twisted vacuum in $\mathcal{H}_a$ -- denoted by $|1 \rangle_a$ -- which is the defect scalar primary in $\mathcal{H}_a$ with the lowest conformal dimension $h_0^a$.

We shall employ a simplification from here on, that is to choose same boundary conditions in \eqref{asymp_4}, at both the entangling points (see Fig. \ref{conf_W} in SM) -- implying, $\widetilde{|\beta_1\rangle}=\widetilde{|\beta_2\rangle}$.


Now using \eqref{srre_ee} and \eqref{asymp_4}, we can compute $\left(\mathcal{Z}^a[q,(c,d)]\right)^n$ and $\mathcal{Z}^a[q^n,(c,d)]$ -- for the replication $q\to q^n$. From this, using \eqref{srre_ee}, we get for the twisted sector SREE as,
\begin{align}
    \cS^a[q,(c,d)] &= \frac{c-24h_0^a}{3}\log\frac{\ell}{\epsilon}
    + \log\left[\frac{N_{ad}^c \, d_c d_d}{d_a|\cC|}\left(g^a_{\beta}\right)^2\right], \label{vN_n_T}
\end{align}
\eqref{vN_n_T} is kind of a correspondence between ASRDoS as in Eq.(3.42) or Eq.(4.24) of \cite{Lin:2022dhv} and twisted CaT-SREE. The appearance of $h_0^a$ in \eqref{vN_n_T} is reminiscent of the fact that we are dealing with defect Hilbert spaces where the corresponding irreps are labelled by the eignevalues of the $T$-matrix which happen to be the conformal dimensions \cite{Harvey:2018rdc, Lin:2022dhv}. We will comment more on this in \cite{DMS:upcoming}. Note that, \eqref{vN_n_T} with $a=\id$ implies $c=d$. This now reproduces the main result of \cite{Saura-Bastida:2024yye} that is Eq.(34).

\textit{Generalized Anyonic Sectors and SymTFT}\textemdash
For $(1+1)$-d CFTs with non-invertible symmetries, the notion of charged sectors arises by inserting TDLs in the time direction and then decomposing the resulting  Hilbert space into irreducible representations, each of which corresponds to a generalized charged sector. As mentioned above, TDLs map local fields to local fields \cite{Chang:2018iay} and their action on a twisted defect field is described by the so-called {\it lasso action} \cite{Chang:2018iay, Lin:2022dhv} under which the resulting defect field might belong to a different defect Hilbert space \cite{Lin:2022dhv}. When this is the case, one needs to consider the {\it total} Hilbert space as, $ \cH_{\text{total}} = \bigoplus_{a}\cH_a,\, \, a \in \cC\, $
with $a: \cH_{a\neq\mathds{1}}\to\cH_{b}$ in general. Then $\cH_{\text{total}}$ can be decomposed into various irreps $\mu$ under the action of TDLs given by the so-called {\it tube algebra} of $\cC$ (denoted Tube$(\cC)$). These irreps are referred to as $\mu$-{\it anyonic sectors} and, remarkably, are in one-to-one correspondence with the bulk anyons $\mu$ of the  Drinfeld center $\cD(\cC)$ (quantum double) of the $(2+1)$-d SymTFT \cite{Lin:2022dhv}. Concretely, in the case of two-dimensional RCFTs, $\cD(\cC)$ is given by $\cC\boxtimes\overline{\cC}$ and thus,  a bulk anyon $\mu\in\cD(\cC)$ is given by $\mu=c \otimes \bar{d}$ with $c,d$ denoting the simple objects in $\cC$. For RCFTs with self-conjugate irreps, we further have, $\mu = c\otimes d$.

\textit{Anyonic CaT-SREE}\textemdash
The above defined $\mu$ labels the irreps of $\mathcal{C}$ and are in one to one correspondence with the generalized charges of \cite{Bhardwaj:2023ayw} (see also \cite{Bartsch:2023wvv}). Following \cite{Lin:2022dhv} we define the relevant projectors as, 
\begin{align}\label{Pmu}
  P^{\mu} &\equiv P^{(c,d)} = \sum_a\,  P^{(c,*)}_a\, P^{(*,d)}_a \nonumber\\
  &= \sum_{a,b,b^\prime}\, S_{0c} S^{*}_{bc} S_{0d} S^{*}_{b^\prime d} \raisebox{-22pt}{
\begin{tikzpicture}
\draw [draw=maroon, thick, decoration = {markings, mark=at position 0.8 with {\arrow[scale=1]{stealth}}}, postaction=decorate] (-1,0.5) -- (1,0.5) node[right]{$b$};
\draw [preaction={draw=white,line width=6pt}, draw=maroon, thick, decoration = {markings, mark=at position 0.9 with {\arrow[scale=1]{stealth}}}, postaction=decorate] (0,0) -- (0,1.5) node[above]{$a$};
\draw [preaction={draw=white,line width=6pt}, draw=maroon, thick, decoration = {markings, mark=at position 0.8 with {\arrow[scale=1]{stealth}}}, postaction=decorate] (-1,1) -- (1,1) node[right]{$b^\prime$};
\end{tikzpicture}
}
\end{align}
This amounts to saying that the charged moments we are seeking for are given by
\begin{align}\label{Zmu}
  \cZ[q,\mu] \equiv \cZ[q,(c,d)] = \sum_{a \in \mathcal{C}}\, \cZ^{a}[q,(c,d)] .
\end{align}
In the language of the anyonic sectors we can rewrite \eqref{asymp_4} -- with equal boundary conditions at both the entangling points -- as,
\begin{align}\label{mu_m}
    \cZ^{a}[q,\mu] = \langle \mu, a \rangle \, \frac{d_\mu}{|\mathcal{\cC}|}\, \frac{\left(g_{\beta}^a\right)^2}{d_a}\,  \tilde{q}^{h^{a}_0 - c/24}\,
\end{align}
where $d_\mu$ denotes the dimension of the bulk anyon and we have used the relation (see Eq.(4.47) of \cite{Lin:2022dhv}): $d_\mu\, \langle \mu, a \rangle =  N_{ca}^{d}\,  d_c\, d_d$. Now $\langle \mu, a \rangle$ is the projection of the bulk anyon $\mu$ onto the boundary line: $a$ \cite{Lin:2022dhv}. Furthermore, to get \eqref{mu_m} from \eqref{asymp_4} we also require: $N^{c}_{ad}=N^{d}_{ca}$ which is true for self-conjugate irreps (see SM for a simple argument).

Using above we can write \eqref{Zmu} as,
\begin{align}
  \cZ[q,\mu] &= \frac{d_\mu}{|\mathcal{C}|}\, \sum_{a \in \mathcal{C}}\, \frac{\langle \mu,a\rangle}{d_a}\, \left(g_{\beta}^a\right)^2 \, \tilde{q}^{h^{a}_0 - c/24}, \nonumber\\
  &=\frac{d_\mu}{|\cC|}\left[\underbrace{\langle \mu,\mathds{1}\rangle \, g_\beta^2 \, \tilde{q}^{\left(-\frac{c}{24}\right)}}_{\text{leading in }\tilde{q}\to 0} + \underbrace{\sum\limits_{a\neq\mathds{1}}\frac{\left(g_\beta^a\right)^2}{d_a}\tilde{q}^{h_0^a-\frac{c}{24}}}_{\text{sub-leading in }\tilde{q}\to 0}\right], \nonumber\\
  &\overset{\tilde{q}\to 0}{\sim} \, \, \frac{d_\mu}{|\cC|} \, \langle \mu,\mathds{1}\rangle \, g_\beta^2 \, \tilde{q}^{\left(-\frac{c}{24}\right)} = \frac{d_\mu}{|\cC|} \, g_\beta^2 \, \tilde{q}^{\left(-\frac{c}{24}\right)}, \label{Zmu_b}
\end{align}
where the last equality is obtained in the following way: we have: $\langle \mu,a\rangle = N^{d}_{ca}$ (see \cite{Lin:2022dhv}). This implies: $\langle \mu,\mathds{1}\rangle = N^{d}_{\mathds{1}c}$. Now: $c\times \mathds{1} = \sum\limits_d \, N^d_{c\mathds{1}} \, d = N^c_{\mathds{1}c} \, c = c$, implying, $\langle \mu,\mathds{1}\rangle=1$.

\textit{Breaking of entanglement equipartition} \textemdash
Now we state the important observation that, w.r.t to the anyonic sectors, {\it entanglement equipartition breaks} -- at the sub-leading order in the UV cut-off $\epsilon$  in the case of categories with non invertible TDLs as opposed to categories with only invertible TDLs. 

We can consider the replication $q\to q^n$ and the $n^{\text{th}}$ power of $\cZ[q,\mu]$ as in \eqref{Zmu_b}. From which, we obtain the following SREE for the anyonic sectors (using \eqref{srre_ee} and noting that: $\tilde{q} = e^{-2W} \sim e^{-4\log\frac{\ell}{\epsilon}}$),
\begin{align}
    &\cS[q,\mu] = \frac{c}{3}\log\frac{\ell}{\epsilon} + \log\frac{d_\mu}{|\cC|} + 2\log g_\beta + \cO\left(\frac{\epsilon}{\ell}\right). \label{main_anyon_res}
\end{align}
\eqref{main_anyon_res} is kind of a correspondence between ASRDoS as in Eq.(4.23) of \cite{Lin:2022dhv} and anyonic CaT-SREE.

We note that if $\cC$ contains only invertible TDLs \footnote{where $\cC$ can be a general symmetry category -- not necessarily a fusion category obtained from an MFC -- like $\mathbb{Z}_2$ whose $\cD(\mathbb{Z}_2)$ is the {\it toric code} with group-like fusion rule given by $\mathbb{Z}_2\times \mathbb{Z}_2$} then $\cD(\cC)$ will contain bulk anyons $\mu$ with $d_{\mu}=1\, \forall \mu$, by construction. Then, it follows that entanglement equipartition holds even at sub-leading order: $\cO(1)$ from \eqref{main_anyon_res}.

However, if $\cC$ contains atleast one non-invertible line, then at least one generalized irrep $\mu$ has $d_\mu > 1$ and we observe a {\it breaking of entanglement equipartition} at sub-leading order.

\Comment{
That is, if we take the difference between two different anyonic sectors \eqref{main_anyon_res},
\begin{align}
    \cS[q,\mu] - \cS[q,\nu] &= \log\frac{d_\mu}{d_\nu} + \cO\left(\frac{\epsilon}{\ell}\right) \label{ent_diff}
\end{align}
one sees that the difference is non-vanishing at $\cO\left(\frac{\epsilon}{\ell}\right)$ if the given category has only invertible TDLs. However, it is non-vanishing at $\cO(1)$ if it has at least one non-invertible TDL. This is a quantifiable distinguishing feature between invertible and non-invertible symmetries.
}

\textit{Discussion and Outlook}\textemdash
We have provided a full symmetry resolution for the entanglement entropy of an interval in a $(1+1)$-d RCFT with a non-invertible symmetry defined through a modular fusion category (MFC). To this end, we used suitable representations and, therefore, suitable notions of generalized charged sectors of the MFC which are given by the anyons of the associated $(2+1)$-d SymTFT. Using this construction, we have defined the projected partition functions in such charged sectors that define the symmetry resolution of the entanglement entropy.

The main assumptions used in this work are: we have a diagonal RCFT with a unique vacuum in the bulk Hilbert space. Furthermore, at times, we have assumed that all of its irreps are self-conjugate. It seems like the diagonal and self-conjugation conditions can be relaxed to allow for the application of the anyonic CaT-SREE to more general scenarios. For instance, the generalized Cardy constraints have been discussed recently in \cite{Gu:2023yhm} in the context of non-diagonal RCFTs and one can use this to construct anyonic CaT-SREE for them. Another generalization is to use the anyonic CaT-SREE construction for arbitrary fusion categories -- e.g., CFT with Haagerup symmetry, as discussed in \cite{Lin:2022dhv}. Another interesting direction is to compute anyonic CaT-SREE for non-unitary theories like the Lee-Yang minimal model: $\cM(5,2)$ -- which is one of the computations in an upcoming work \cite{DMS:upcoming}.\\

\noindent\textit{Note added}\textemdash
During the final stages of preparation of this manuscript, we were made aware of partly related developments in \cite{Choi:2024tri, Choi:2024wfm, Bhardwaj:2024igy, Heymann:2024vvf, GarciaEtxebarria:2024jfv}.\\

\noindent\textit{Acknowledgments}\textemdash
We express our sincerest gratitude to Christian Northe, Daniel Grumiller, Ho Tat Lam, I\~{n}aki García-Extebarria, Jishu Das, Mathew Bullimore, Michele Fossati, Nabil Iqbal, Naveen Balaji Umasankar, Rajath Radhakrishnan, Saranesh Prembabu, Shovon Biswas, Statis Vitouladitis, Thomas Bartsch, Timmavajjula Venkata Karthik, and Tin Sulejmanpasic for many insightful discussions. We thank Brandon Rayhaun, Yichul Choi, Yunqin Zheng, Christian Copetti, Daniel Pajer, Lakshya Bhardwaj, Sakura Schafer-Nameki, Jared Heymann, Thomas Quella, Angel Uranga, Jes\'{u}s Huertas and I\~{n}aki Garc\'{i}a Etxebarria for agreeing to coordinate the release of partially related upcoming works with us. The work of A.D. is supported by the STFC Consolidated Grant ST/T000600/1 -- ``Particle Theory at the Higgs Centre''.  J.M.-V. thank the financial support of Spanish Ministerio de Ciencia e Innovación PID2021-125700NAC22.  The work of P.S.-B. is supported by Fundaci\'on S\'eneca, Agencia de Ciencia y Tecnolog\'ia de la Regi\'on de Murcia, grant 21609/FPI/21.

\bibliographystyle{utphys}
\bibliography{prl_v2}

\onecolumngrid

\appendix

\section{Suppplemental Material}

\subsection{Quantum dimension and a property of S-matrix}
An important property about TDL that can be obtained from \eqref{tdl_local} is its {\it quantum dimension} which can be defined as its eigenvalue when it acts on the vacuum $|1\rangle$ in \eqref{tdl_local}, that is,
\begin{align}
    b |1\rangle = \frac{S_{b1}}{S_{11}}|1\rangle \equiv d_b|1\rangle, \label{q_dim}
\end{align}
and so $d_b \equiv \langle 1|b|1\rangle = \langle b\rangle$. Now, using the one-to-one correspondence between TDLs and bulk primaries one can define the quantum dimension of a bulk primary labelled as $|\phi_a\rangle$ as $d_a\equiv \frac{S_{1a}}{S_{11}}$ whose physical meaning can be inferred as the ratio of the asymptotic size of the $\cV_a$ module over the identity module $\cV_\mathds{1}$ (see \cite{Northe:2023khz} for more details). It can be shown that $d_a\geq 0$ for any unitary CFT with a unique vacuum \cite{Chang:2018iay, proto_Simon}.

A line $b$ is invertible iff $b^n=\id$ with $n\in\mathbb{N}$ which further implies that its quantum dimension is $d_b=1$. On the contrary, $b$ is a non-invertible line iff $b^2=\mathds{1} + \ldots$, which thereby implies that $d_b>1$. To see this note that the quantum dimensions furnish a representation for the fusion coefficients (see \cite{proto_Simon}),
\begin{align}\label{non-inv_rep}
    d_a \, d_b = \sum\limits_{c}N^c_{ab} \, d_c,
\end{align}
implying: $N^{1}_{aa} = 1$. Thus, from Eq.(\ref{non-inv_rep}), we get:
\begin{align}
    d_a^2 = 1 + \sum\limits_{c\neq 1}N^c_{aa}d_c \implies d_a>1. \label{qDim_non-inv}
\end{align}

Below we present a nice property of the $S$-matrix (see (17.7) of \cite{proto_Simon}) which shall be very useful,
\begin{align}
    S_{ab} = S_{ba} = S_{\bar{a}\bar{b}} = S_{\bar{b}\bar{a}} = S^{*}_{\bar{a}b} = S^{*}_{b\bar{a}} = S^{*}_{a\bar{b}} = S^{*}_{\bar{b}a}. \label{S_prop} 
\end{align}

\subsection{BCFT approach to EE}
Consider an interval $A$ of length $\ell$ in a CFT. We know that its entanglement entropy - at leading order - is given as:
\begin{align}
    \cS_A = \frac{c}{3}\log\left(\frac{\ell}{\epsilon}\right) + \cO(1), \label{EE}
\end{align}
where $\epsilon$ denotes the UV-cutoff. This result dates back to  \cite{Cardy:1988tk} (see also \cite{Calabrese:2004eu}). The primary ways to obtain this quantity was to consider the replicated partition function and then construct R\'{e}nyi entropies from it which when analytically continued to a particular limit $n\to 1$ yields the von Neumann entropy which when performed upon an interval gives its entanglement entropy. There is also the twist field approach to obtaining the above entanglement entropy (see \cite{Calabrese:2009qy} for review). However, another elegant way to obtain this result is via the boundary CFT approach of \cite{Cardy:2004hm} (see \cite{DiGiulio:2022jjd, Kusuki:2023bsp, Saura-Bastida:2024yye} for reviews). For sake of completeness, we briefly review this approach below \cite{Ohmori:2014eia}.

\begin{figure}[h]
\centering
\includegraphics[width=0.75\textwidth]{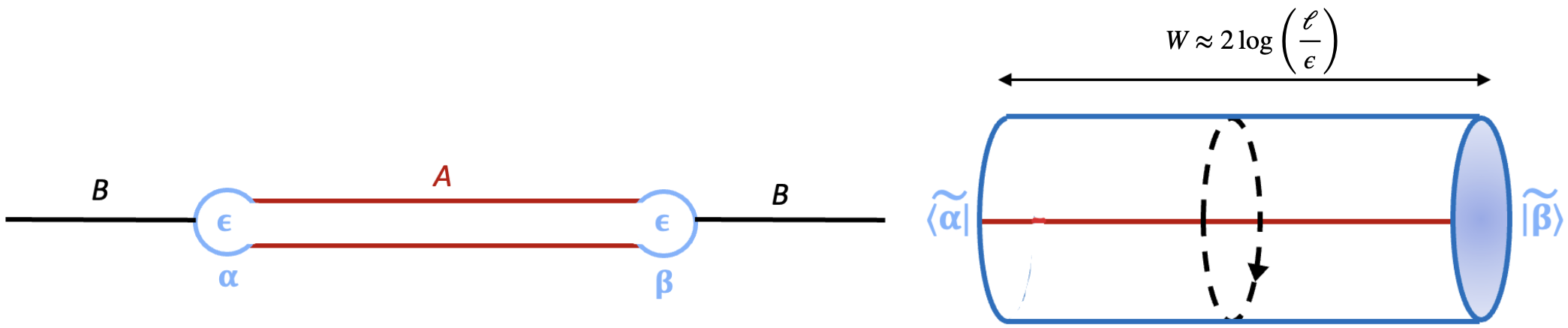}
\caption{\emph{\small The factorization $\alpha\beta$ imposes disks $\epsilon \ll 1$ with boundary conditions $\alpha$ and $\beta$ (left figure). The resulting manifold is replicated and after tracing over $\mathcal{H}_{B,\beta\alpha}$\,, a conformal transformation yields an annulus of width $W=2\log\left(\frac{\ell}{\epsilon}\right) + \mathcal{O}(\epsilon)$ and circumference of $2\pi n$}}
\label{conf_W}
\end{figure}

The interval $A$ entangling points between region $A$ and its complement $B$ are labelled by boundary states $\widetilde{|\alpha\rangle}$ and $\widetilde{|\beta\rangle}$. To take care of the UV divergences at the entangling points small disks of radius $\epsilon$ is introduced which is also the UV cutoff here (see figure Fig.\ref{conf_W}). These need to be conformal boundary states since we demand conformal invariance. Then the idea is to replicate this manifold and then conformally map the resulting manifold to an annulus of width $W$. This width: $W=2\log\left(\frac{\ell}{\epsilon}\right) + \mathcal{O}(\epsilon)$, is related to the torus modular parameter by the conformal map: $\tau = \frac{\text{i}\pi}{W}$. Thus, we have $q=e^{-\frac{2\pi^2}{W}}$ and $\tilde{q}=e^{-2W}$. 

In this geometry, we have the following factorisation of the bulk Hilbert space: $\cH = \cH_{A,\alpha\beta}\otimes \cH_{B,\beta\alpha}$. The reduced density matrix of region $A$ is written as:
\begin{align}
    \rho_A = \frac{q^{L_0 - \frac{c}{24}}}{Z(q)}, \label{partn_fn_neq}
\end{align}
where $Z(q)=\tr(q^{L_0-\frac{c}{24}})$ is the {\it open string} partition function -- with $q=e^{-2\pi^2/W}$ -- and the trace is subjected to the boundary conditions $\widetilde{|\alpha\rangle}$ and $\widetilde{|\beta\rangle}$. From this one can compute the R\'{e}nyi entropy and subsequently the von Neumann entropy in the usual way \cite{Cardy:1989ir, Cardy:2016fqc}:
\begin{align}
    &\tr(\rho_A^n) = \frac{Z(q^n)}{(Z(q))^n}, \nonumber\\
    &\cS_n(q) = \frac{1}{1-n}\log\frac{Z(q^n)}{(Z(q))^n}, \qquad \cS_{\text{vN}} = \lim\limits_{n\to 1}\cS_n(q). \label{vN}
\end{align}

If we perform an $S$-transformation then the {\it open string} partition function -- which is a trace over {\it open string states} -- can be written as an amplitude of propagation of {\it closed string states} as follows,
\begin{align}
    Z(q) &= \sum_i \, n^i_{\alpha\beta} \chi_i(q) = \widetilde{\langle \alpha|}\tilde{q}^{L_0-\frac{c}{24}}\widetilde{|\beta\rangle}, \qquad Z(q^n) = \sum_i \, n^i_{\alpha\beta} \chi_i(q^n) = \widetilde{\langle \alpha|}\tilde{q}^{\frac{1}{n}\left(L_0-\frac{c}{24}\right)}\widetilde{|\beta\rangle}, \label{open_close_neq}
\end{align}
where $\chi(q)$s are the open string characters with $n^i_{\alpha\beta}$ denoting their multiplicities, $\widetilde{|\alpha\rangle}$ and $\widetilde{|\beta\rangle}$ are conformally invariant boundary states in the {\it closed string} spectrum, also referred to as {\it Cardy states}. They are expressed as linear combination of the so called {\it Ishibashi states} $|j\rangle\rangle$ \cite{Cardy:1989ir},
\begin{align}
    \widetilde{|\alpha\rangle} = \sum_j\frac{S_{aj}}{\sqrt{S_{1j}}}|j\rangle\rangle, \label{Cardy_Ishibashi}
\end{align}
Let us now evaluate the closed string amplitude as given in \eqref{open_close_neq} in the limit of: $\epsilon\to 0$ or, $\tilde{q}\to 0$  (or, $q\to 1$),
\begin{align}
    Z(q^n) = \widetilde{\langle \alpha|}\tilde{q}^{\frac{1}{n}\left(L_0-\frac{c}{24}\right)}\widetilde{|\beta\rangle} \, \, \overset{\tilde{q}\to 0}{\underset{\epsilon\to 0}{\sim}} \, \, \widetilde{\langle \alpha|}1\rangle \langle 1\widetilde{|\beta\rangle}\tilde{q}^{\frac{1}{n}\left(h^\id_{0}-\frac{c}{24}\right)} \sim e^{g^{*}_{\alpha}}e^{g_\beta}\tilde{q}^{\frac{1}{n}\left(-\frac{c}{24}\right)}, \label{g-fn_neq}
\end{align}
where $g_\alpha\equiv \log\langle 1\widetilde{|\alpha\rangle}$ is the {\it Affleck-Ludwig} boundary entropy or the so called $g$-function \cite{Affleck:1991tk, Friedan:2003yc} and $|1\rangle$ denotes the vacuum state of the bulk Hilbert space with scaling dimension $h^\id_{0}=0$. Now the R\'{e}nyi entropy becomes (from \eqref{vN}),
\begin{align}
    \cS_n(q) &\sim \frac{1}{1-n}\left[\log\frac{\tilde{q}^{\frac{1}{n}\left(-\frac{c}{24}\right)}}{\tilde{q}^{n\left(-\frac{c}{24}\right)}} + (1-n)(g_\alpha^{*} + g_\beta)\right]\sim\frac{1}{1-n}\left[\frac{c}{6}\left(\frac{1}{n}-n\right)\log\frac{\ell}{\epsilon} + (1-n)(g_\alpha^{*} + g_\beta)\right], \label{EE_1}
\end{align}
which leads to the entanglement entropy as,
\begin{align}
    \cS_{A}(q) = \lim\limits_{n\to 1}S_n(q) = \frac{c}{3}\log\frac{\ell}{\epsilon} + g_\alpha^{*} + g_\beta + \cO\left(\frac{\epsilon}{\ell}\right), \label{EE_2}
\end{align}
where the $g$-function terms above are sub-leading terms at $\cO(1)$. Thus, the above reproduces the entanglement entropy as given in \eqref{EE}.

\subsection{Untwisted CaT-SREE}
In this section we briefly overview the construction of \cite{Saura-Bastida:2024yye}. Note that, in general TDLs (or, for simplicity, say Verlinde lines) are non-invertible and their fusion is described by a fusion algebra which is generically non-group like. In this case, owing to the chiral symmetry of the chiral algebra $\cA$ of the CFT, we have the the bulk Hilbert space decomposes as given in \eqref{H1}. Hence, the corresponding projectors to these individual Verma modules can be defined as \cite{Lin:2022dhv},
\begin{equation}
P_\mathds{1}^{(a,a)}\equiv\sum_{b\in\cC}S_{1a}S^{*}_{ba} 
\raisebox{-5pt}{
\begin{tikzpicture}
\draw [draw=maroon, thick, decoration = {markings, mark=at position 0.6 with {\arrow[scale=1]{stealth}}}, postaction=decorate] (-1,0) -- (1,0) node[right]{$b$} ;
\end{tikzpicture}
}
\equiv \frac{d_a}{|\cC|}\sum_{b\in\cC}\chi_a^{*}(b)
\raisebox{-5pt}{
\begin{tikzpicture}
\draw [draw=maroon, thick, decoration = {markings, mark=at position 0.6 with {\arrow[scale=1]{stealth}}}, postaction=decorate] (-1,0) -- (1,0) node[right]{$b$};
\end{tikzpicture}
}
\label{RCFT-untwisted-P}
\end{equation}
where $d_a$ is the quantum dimension of the line $a$, $|\cC|\equiv\sum\limits_{b\in\cC}d_b^2$ is the total quantum dimension of the modular fusion category (MFC) \cite{Bhardwaj:2017xup} -- which equals the square of the total quantum order of the MFC $\cC$ \cite{rowell2009classification}. It can be shown that for an RCFT, with only self-conjugate irreps, $|\cC|=\frac{1}{S_{11}^2}$ (see appendix of \cite{Northe:2023khz} for a proof). $\chi_a^{*}(b)\equiv\frac{S^{*}_{ba}}{S_{11}}$ which reduces to: $\chi_a^{*}(\mathds{1})\equiv\frac{S^{*}_{1a}}{S_{11}}=d_a$ (using \eqref{S_prop}) -- just like in the group-like case, we had: $\chi_r^{*}(e)=d_r$, which gives the dimension of the irrep $r$. The second equality above helps to write the projectors above in a compact way that has a resemblance with the projectors to various irreps of the group like case as in \eqref{eq:proj_Pi_r} \cite{Saura-Bastida:2024yye}. 

The above projectors give the projections $P_\mathds{1}^{(a,a)}: \cH_1\to \cV_a\otimes\overline{\cV_a}$\,. Furthermore, they can be shown to satisfy the defining projector properties of idempotency and completeness \cite{Lin:2022dhv}. We can now define the untwisted projected partition function as,
\begin{align}
\tr \left[{P_\mathds{1}}^{(a,a)} \rho\right] &\equiv \tr_{\cH_{\mathds{1}}^{(a,a)}}\, q^{L_0 - c/24} \, \bar{q}^{\bar{L}_0-\bar{c}/24} =\sum_{b}S_{1a}{S^{*}_{ba}}
\raisebox{-11pt}{
\begin{tikzpicture}
\draw [draw=maroon, thick, decoration = {markings, mark=at position 0.6 with {\arrow[scale=1]{stealth}}}, postaction=decorate] (-1,0.5) -- (1,0.5) node[right]{$b$};
\draw (-1,0) rectangle (1,1);
\end{tikzpicture}
} \nonumber\\
&=\sum_{b,c}S_{1a}S^*_{ba}\frac{S_{bc}}{S_{1c}}\chi_c(q)\overline{\chi_c(q)}=\chi_a(q)\overline{\chi_a(q)}=\tr_{\cV_a\otimes\overline{\cV_a}}q^{L_0 - c/24} \, \bar{q}^{\bar{L}_0-\bar{c}/24}.
\label{eq:RCFT-untwisted-P-proof}
\end{align}

Now the computation paralleling the group-like case can be employed (see \cite{Kusuki:2023bsp}), to compute the respective charged moments and thereby the projected partition functions w.r.t. the projectors given in \eqref{RCFT-untwisted-P} with the following replacement: 
\begin{align}
    \cZ[q^n,(a,a)] = \frac{d_a}{|\cC|}\sum_{b\in \cC}\chi^{*}_a(b)\frac{Z[q^n,(b,b)]}{(Z[q])^n}. \label{charged_srre_eval}
\end{align}

However, we would need to make another replacement. In the group like case, one needed $G$-invariant (or, symmetric) Cardy states to define the charged moment $Z[q^n,e]$ -- with $e$ being the identity group element. Similarly, here we would require the notion of {\it categorical invariant} boundary states which would now be invariant under the action of TDLs belonging to a modular fusion category. Such notions were introduced and discussed in \cite{Choi:2023xjw} to which we now turn or attention to.

Following \cite{Choi:2023xjw} we define two notions of symmetric boundary states w.r.t. the fusion category in question. 

\subsubsection{Categorical symmetric boundary states}
Given a modular fusion category $\cC$, a conformal boundary state $\widetilde{|\beta\rangle}$ is called $\cC$-weakly symmetric if,
\begin{align}
    a \widetilde{|\beta\rangle} = \widetilde{|\beta\rangle} + \ldots, \quad \forall \, \, a\in\cC, \label{weak_C}
\end{align}
and $\widetilde{|\beta\rangle}$ is called $\cC$-strongly symmetric if,
\begin{align}
    a \widetilde{|\beta\rangle} = d_a\widetilde{|\beta\rangle}, \quad \forall \, \, a\in\cC, \label{strong_C}
\end{align}
where $\cC$-weakly symmetric notion ensures {\it topological endability} of the TDLs at the boundary. For a connection of these two symmetric notions with {\it categorical anomalies} see \cite{Choi:2023xjw}.

Note that, if a conformal boundary state is strongly symmetric then it is weakly symmetric too but the converse isn't true. Furthermore, for the group like case, the notion of strongly symmetric conformal boundary states reduces to the notion of symmetric (or, $G$-invariant) conformal boundary states (since for invertible TDLs: $d_a=1$, see SM for a proof).

So, if a fusion category admits any of the above two notions of categorical symmetric Cardy states then we can symmetry resolve w.r.t. that category of TDLs \cite{Saura-Bastida:2024yye}. With this, the algorithm of computing the symmetry resolved entanglement entropy mimics the group like case. For the projected partition functions we get -- in the limit of $q\to 1$, at the leading order (since, $\chi_a^{*}(\mathds{1})=d_a$),
\begin{align}
    &\mathcal{Z}^{\mathds{1}}[q^n,(a,a)] = \frac{d_a}{|\cC|}\sum_{b\in \cC}\chi_a^*(b)\frac{Z(q^n,b)}{(Z(q))^n} \sim \frac{d^2_a}{|\cC|}\frac{Z(q^n)}{(Z(q))^n}, \qquad
    \left(\mathcal{Z}^{\mathds{1}}[q,(a,a)]\right)^n \sim \left(\frac{d_a^2}{|\cC|}\right)^n ,\label{eq:gen_finite_cat}
\end{align}
which gives the following expressions for the symmetry resolved entanglement entropy:
\begin{align}
    &\cS_{n}^{\mathds{1}}[q,(a,a)] = \frac{1}{1-n}\left[\frac{c}{6}\left(\frac{1}{n}-n\right)\log\frac{\ell}{\epsilon} + (1-n)(g_\alpha^{*} + g_\beta) + (1-n)\log\frac{d_a^2}{|\cC|}\right] + \cO\left(\frac{\epsilon}{\ell}\right), \nonumber\\ &\cS_A^{\mathds{1}}[q,(a,a)] = \frac{c}{3}\log\frac{\ell}{\epsilon} + (g_\alpha^{*} + g_\beta) + \log\frac{d_a^2}{|\cC|} + \cO\left(\frac{\epsilon}{\ell}\right), \label{srre_Cat_uT}
\end{align}
where the label -- $(a,a)$ -- denotes the irrep (or, charged sector) in this case, that is, it labels the Verma module: $\cV_a\otimes\overline{\cV_a}$. Furthermore, $g_\beta\equiv\log\langle 1\widetilde{|\beta\rangle}$ with $\widetilde{|\beta\rangle}$ being either a strongly or weakly symmetric Cardy state w.r.t. the category $\cC$.

Note that, if $a$ is an invertible TDL then, entanglement equipartition holds even at sub-leading order in the UV cut-off -- $\cO(1)$ else if it is non-invertible then equipartition only holds at leading order, as can be readily seen from \eqref{srre_Cat_uT}.

\subsection{From torus to annulus}
For the twisted sector case, note that the action of TDLs shuffles defect fields among each other living in different defect spaces (or twisted sectors). So, one needs to be careful while constructing boundary states in this case. The reason being the corresponding twisted Cardy states will also get similarly shuffled. A way to circumvent this is to consider the following reasonable approximation.

We learn from the untwisted case that, the Cardy states form non-negative integer matrix representations -- \textit{NIM-reps} -- under the action of TDLs \cite{Choi:2023xjw}. Infact for a diagonal RCFT, the \textit{boundary fusion algebra} is the same as the bulk fusion algebra. 

\begin{align}
    a\widetilde{|b\rangle} = \sum\limits_{c\in\cC}N^c_{ab} \, \widetilde{|c\rangle}, \label{bdry_bulk_fuse}
\end{align}
where $N^c_{ab}$ are the bulk fusion coefficients and $\widetilde{|b\rangle}$ denotes the untwisted Cardy state -- here labelled by the label of the corresponding primary.

Hence, the idea is to start with a twisted sector and construct projectors of it as in \cite{Lin:2022dhv} for the torus partition function. Then, we consider the assumption that there exists a unique vacuum in the untwisted sector whose contribution dominates in the high temperature limit: $\ell>>\epsilon$. This is the approximation of $b^{\prime\prime}\to\mathds{1}$ in \eqref{eval_F} and in \cite{Lin:2022dhv} (see their appendix for more details).

After this, we land in the untwisted sector with a TDL insertion along the spatial $S^1$ as in \eqref{torus_4}. Now, we can introduce the interval $A$ here and consider Cardy states at its entangling points. As before, here the untwisted Cardy states will form NIM-reps and hence will not be mapped out of the untwisted sector under the action of TDLs. Therefore, we can use the bulk fusion rule as the \textit{boundary fusion rule} here.

In this way, we can reach the annulus partition function from the torus partition function and rest all calculation of the symmetry resolution -- either for the group-like case or the non-invertible case -- follows from here. This way we can get twisted CaT-SREE results and connect them to the twisted sector analogue of ASRDoS of \cite{Lin:2022dhv}. We comment on this more in \cite{DMS:upcoming}.

However, if we demand to have the interval $A$ inserted in the torus partition function in the twisted sector as in \eqref{eval_F}, then we need to consider action of TDLs on {\it twisted Cardy states} where these will be irreps of some kind of \textit{boundary fusion algebra} along the lines as discussed in \cite{Kitaev:2012, Jia:2020jht, Bullivant:2020xhy, delaFuente:2023whm, Ishikawa:2005ea, Kojita:2016jwe, Konechny:2019wff, Konechny:2024ixa}. We defer such computations to the future \cite{DMS:upcoming}.

\subsection{Fusion of lines $b$ and $b^\prime$}
To evaluate \eqref{eval_F} consider fusing $b$ and ${b^\prime}$ by inserting an orthonormal set of associators: $\{\psi_j\}_j$ of $\mathrm{Hom}(b^{\prime\prime},b\otimes b^\prime)$, to find
 -- in the limit of $b^{\prime\prime}\to\mathds{1}$, which in turn amounts to assuming that there exists a unique vacuum in the theory (see \cite{Lin:2022dhv} for more details),
\begin{align}
    &\tr \left[P_a^{(c,d)}\rho\right] \, \overset{b^{\prime\prime}\to\mathds{1}}{\sim} \, S_{1c}S_{1d} \, d_a\left(\sum\limits_b\frac{S^*_{bc}S^*_{\bar{b}d}S_{ba}}{S_{1a}}\frac{1}{d_b}\right) \, \frac{1}{d_a} \, 
    \vcenter{\hbox{\begin{tikzpicture}
        \draw (-1,-.5) rectangle (1,.5);
        \draw [draw=maroon, thick, decoration = {markings, mark=at position 0.6 with {\arrow[scale=1]{stealth}}}, postaction=decorate]  (1,0) -- node(A)[below] {$_a$} (-1,0);
        \end{tikzpicture}}} = \, S_{1c}S_{1d} \, d_a\left(\sum\limits_b\frac{S^*_{bc}S_{bd}S_{ba}}{S_{1a}}\frac{1}{d_b}\right) \, \frac{1}{d_a} \,
    \vcenter{\hbox{\begin{tikzpicture}
        \draw (-1,-.5) rectangle (1,.5);
        \draw [draw=maroon, thick, decoration = {markings, mark=at position 0.6 with {\arrow[scale=1]{stealth}}}, postaction=decorate]  (1,0) -- node(A)[below] {$_a$} (-1,0);
        \end{tikzpicture}}} \nonumber\\   
    &= \, S_{1c}S_{1d} \, \left(\sum\limits_b\frac{S^*_{bc}S_{bd}S_{ba}}{S_{1b}}\right) \, \frac{1}{d_a} \, 
    \vcenter{\hbox{\begin{tikzpicture}
        \draw (-1,-.5) rectangle (1,.5);
        \draw [draw=maroon, thick, decoration = {markings, mark=at position 0.6 with {\arrow[scale=1]{stealth}}}, postaction=decorate]  (1,0) -- node(A)[below] {$_a$} (-1,0);
        \end{tikzpicture}}} 
    = \, S_{1c}S_{1d} \, \frac{N^c_{ad}}{d_a} \,
        \vcenter{\hbox{\begin{tikzpicture}
        \draw (-1,-.5) rectangle (1,.5);
        \draw [draw=maroon, thick, decoration = {markings, mark=at position 0.6 with {\arrow[scale=1]{stealth}}}, postaction=decorate]  (1,0) -- node(A)[below] {$_a$} (-1,0);
        \end{tikzpicture}}} \, \, \, \overset{\text{under }S^{-1}}{=} \, \, \, S_{1c}S_{1d} \, \frac{N^c_{ad}}{d_a} \,
    \vcenter{\hbox{\begin{tikzpicture}
        \draw (-.5,-1) rectangle (.5,1);
        \draw [draw=maroon, thick, decoration = {markings, mark=at position 0.6 with {\arrow[scale=1]{stealth}}}, postaction=decorate]  (0,-1) -- node(A)[right] {$_a$} (0,1);
        \end{tikzpicture}}}    
    \label{asymp_3}
\end{align} 
where the second equality in the first line comes from noting that $S^{*}_{\bar{b}d}=S_{bd}$ from \eqref{S_prop}.

\subsection{Importance of non-invertible TDLs for symmetry resolution}
Consider a diagonal RCFT -- not necessarily with all irreps being self-conjugate -- with an MFC that contains only invertible lines: $\cC_{\text{inv}}=\lbrace \id, a_1, a_2,\ldots,a_n\rbrace$ such that $a_i^{m_i}=\mathds{1}$ with $m_i\in\mathbb{N}$, $\forall \, \, i\in\lbrace1, \ldots, n\rbrace$. This leads to: $\bar{a}_i=a_i^{m_i-1}$. Invertibility of the lines imply, in general:  
\begin{align*}
    a_i \times a_j = a_k,
\end{align*}
where $i,j,k\in\lbrace1, \ldots, n\rbrace$. Now let us prove that we cannot have $a_k=a_j$ or, $a_k=a_i$ for non-trivial lines. To prove the first claim let us proceed by way of contradiction. Then, 
\begin{align*}
    &a_i \times a_j = a_j, \qquad \Rightarrow \, \, a_i \times a_j \times\bar{a}_j = a_i = \id
\end{align*}
which is a contradiction since $a_i\neq\id$. Thus, our above claim is correct. Similarly, we cannot have: $a_j\times a_i = a_i$. Though, we can have: $a_i\times a_j=\mathds{1}$, in which case: $\bar{a}_i=a_j$. So, the only other possibility is:
\begin{align}
    a_i\times a_j = a_k, \label{non-inv_pres}
\end{align}
for $i\neq j\neq k$. Hence, \eqref{non-inv_pres} is the only non-trivial fusion rule of the category other than $a_i \times \bar{a}_i=\mathds{1}$. 

Since, we are in a diagonal theory $a_i\leftrightarrow \widetilde{|a_i\rangle}$ where $\widetilde{|a_i\rangle}$ is a conformal boundary state. Thus, \eqref{non-inv_pres} implies that the action of invertible lines on the Cardy states is as follow:
\begin{align}
    a_i\widetilde{|a_j\rangle} = \widetilde{|a_k\rangle}, \, \, \forall \, \, i,j,k \, \, \text{distinct} \label{non-inv_pres22}
\end{align}
Hence, there does not exist any $\cC_{\text{inv}}$-weakly symmetric boundary state. So, we cannot group-like or category-like SREE either $\cC_{\text{inv}}$ or any of its sub-category. This means, we need atleast one non-invertible line in the MFC (or symmetry category in general) so as to furnish a weakly symmtric state w.r.t. that category or w.r.t. its sub-category. Then, we can go on with consider symmetry resolving w.r.t. them. 

The above explains -- in a way -- why the BCFT approach to symmetry resolution \cite{Kusuki:2023bsp, DiGiulio:2022jjd, Saura-Bastida:2024yye} is more powerful than the twist field approach as given in \cite{Goldstein:2017bua, Calabrese:2021wvi, Benedetti:2024dku}. We will comment more on this in \cite{DMS:upcoming}.

\subsection{Simple argument to show: $N^{d}_{ca}=N^c_{ad}$}
Let us assume, as before, we have an RCFT with self-conjugate irreps. From the Verlinde formula we have \cite{DiFrancesco:1997nk}:
\begin{align}
    N^d_{ac} &= \sum\limits_b\frac{S_{bc}S_{ba}S^*_{bd}}{S_{1b}} = \underbrace{\sum\limits_b\frac{S_{ba}S^*_{b\bar{d}}S_{bc}}{S_{1b}}}_{(\text{as, }d=\bar{d})} = \underbrace{\sum\limits_b\frac{S_{ba}S_{bd}S^*_{b\bar{c}}}{S_{1b}}}_{(\text{as, }S^*_{b\bar{a}}=S_{ba} \, \, \text{from \eqref{S_prop}})} = \underbrace{ \sum\limits_b\frac{S_{ba}S_{bd}S^*_{bc}}{S_{1b}}}_{(\text{as, }\bar{c}=c)} = N^c_{da} = N^c_{ad}. \label{Ver_fuse}
\end{align}

\end{document}